\documentclass[conference]{ieeetran}
\usepackage{ amsmath, amssymb, multicol, graphicx, gensymb}%pupr_paper,
\usepackage{flushend}%\usepackage{Times}

\title{Single Langmuir Probe Diagnostics Device}

\author{\IEEEauthorblockN{Angel Gonz\'alez-Lizardo\IEEEauthorrefmark{1},
Jairo Rond\'on\IEEEauthorrefmark{1}, Felix A. Cuadrado-Rodr\'iguez\IEEEauthorrefmark{2}}%
\bigskip
\IEEEauthorblockA{\IEEEauthorrefmark{1}Polytechnic University of Puerto Rico,\\
 agonzalez@pupr.edu}
\IEEEauthorblockA{\IEEEauthorrefmark{1}Polytechnic University of Puerto Rico,\\
 jrondon@pupr.edu}
\IEEEauthorblockA{\IEEEauthorrefmark{2}Amazon Web Services (AWS),\\
 facrodz@gmail.com}
 }

\begin{document}
\maketitle

\begin{abstract}
Seeking to improve and innovate the technology currently used in their research work, the Polytechnic University of Puerto Rico’s Plasma Laboratory designed and built a portable device able to generate a voltage sweep for an electrostatic probe (namely, a Single Langmuir Probe, or SLP) and to perform the computations necessary to determine the plasma temperature, and density as well as other parameters. The device uses a Raspberry Pi 4 to generate the voltage signal which is amplified through electronic circuitry in the range of -300V to +300V, and applied to a SLP. The device is able to capture the current returning from the SLP to extract the relevant information from the IV characteristic and perform the computations necessary to obtain plasma electron density, plasma potential, floating potential, and  electron temperature. All data is displayed through a touchscreen by a Graphical user interface developed in PyQT5. This device provides continuous measurement of plasma parameters during the realization of experiments at the laboratory.
\end{abstract}

\section{INTRODUCTION}

Plasma is a state of matter where, when a gas is heated to very high temperatures, atoms lose electrons from their outer shells, forming positive ions in a sea of free electrons. This state can also be produced by heating a neutral gas to elevated temperatures, which highlights the importance of studying it \cite{gonzalez2024sterilization}.  %[a]
This work is concerned with the design and development of a portable plasma diagnostics device for electrostatic probes. The device is capable of generating a voltage ramp and capture the current flowing through the same circuit. The concept is implemented using a Single Langmuir Probe (SLP) \cite{opac-b1133704, gurnett2017introduction}, \cite{chen2003langmuir}, but can be easily extended to other probes. A Raspberry Pi is used to control the diagnostic process and as a human machine interface. The Raspberry Pi calculates the various plasma parameters based on the current/voltage characteristic. An power amplifier is used to bring the Raspberry Pi output voltage ramp from the  from 0-4V range to a range of ± 300 V to be applied to the Langmuir Probe. The device is compact in size, enclosed and properly ventilated.

The Diagnostic Device is able to obtain key plasma parameters such as electron temperature, floating, and plasma potential from the SLP, and computes others such as the electron density, the Debye length, and the Larmor Radius \cite{Chen84}. The Diagnostic Device is an important improvement to the diagnostics tools at the Polytechnic University of Puerto Rico Plasma Engineering Laboratory given its smaller size and light weight as opposed to the previous measurement system which included a computer running matlab and driving a Keithley 2400 source meter \cite{KT2002}.

Reducing the size and weight was achieved by designing a new power system to eliminate the Keithley 2400 and adding the Raspberry Pi \cite{RBPI4B} running a Python program which eliminated the use of matlab. Another major contribution is the added Human Machine Interface (HMI) implemented, which provides the user control over the voltage ramp magnitude as well as a visual of the plasma parameters obtained thus eliminating the need for an external computer.

\section{SUBSYSTEMS}

The Diagnostics Device is able to generate a voltage ramp from -300V to 300V while reading the circuit current and performing the calculations for plasma diagnostics. All this in a small portable device. The Device different subsystems and its purpose are explained here. Some of the sub-systems discussed in this section are: Summing Amplifier, High Voltage Amplifier, Current Sensing Amplifier, and Power Supply, etc.

\subsection{Micro-controller – Raspberry Pi}
The Raspberry Pi 4 Model B was chosen a for this device among many micro-controllers in the market, because it is provided with the Raspbian operating system, it has internet capabilities, 8Gb of ram, HDMI ports, open source programming and a wide variety of peripherals available for purchase.

\begin{figure}
  \centering
  \includegraphics[width=\columnwidth]{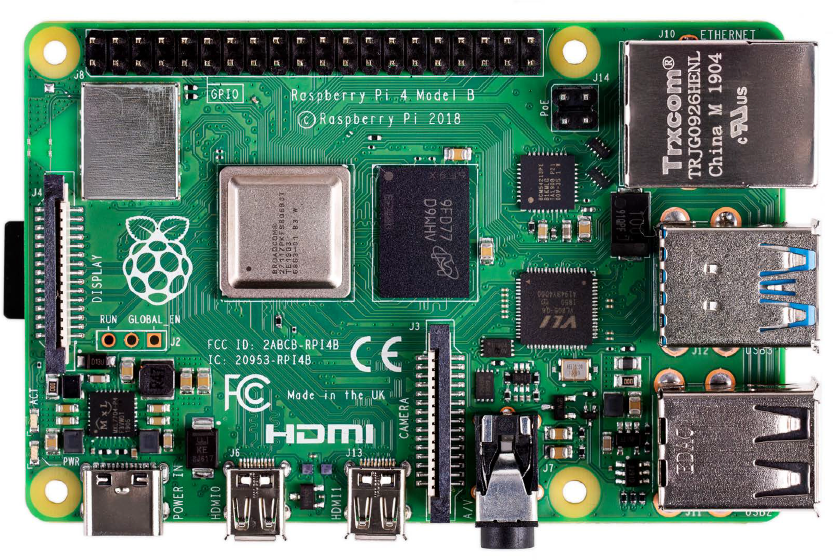}
  \caption{Raspberry Pi 4 Model B \protect\cite{RBPI4B}}\label{rbpi4b}
\end{figure}

The Raspberry Pi shown in Figure \ref{rbpi4b}, counts with various clocks (e.g. ARM, Core, V3D, ISP, H264, HEVC) inside the SoC, monitored by the firmware, and Dynamic Voltage and Frequency Scaling (DVFS). This may result in significant SoC power reductions, hence reducing the overall heat produced. A  stepped CPU governor also allow finer-grained control of ARM core frequencies, increasing the DVFS effectiveness \cite{sedra2020microelectronic}.

\subsection{Touchscreen Display}
\begin{figure}
  \centering
  \includegraphics[width=\columnwidth]{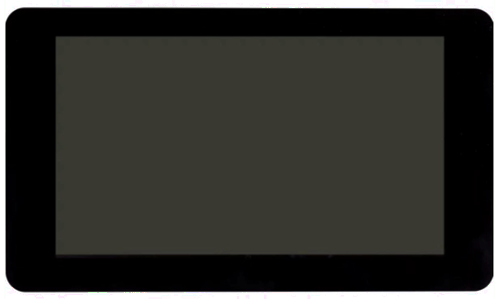}
  \caption{Raspberry Pi Touchscreen\\ (https://www.raspberrypi.org/products/raspberry-pi-touch-display/)}\label{rbpits}
\end{figure}

A Raspberry Pi Touchscreen display is included in the design enable easy user input and control, as well as display of information provided by the system, eliminating the need for an external monitor, keyboard and a mouse.
The Raspberry Pi Touchscreen shown in Figure \ref{rbpits}, is a 7” monitor with an 800 x 480 resolution. Only two connections are needed to connect to the Raspberry Pi. The touchscreen drivers support 10-finger touch and an on-screen keyboard.

\subsection{DAQC2 – Data Acquisition Board}

To- generate the analog voltage ramp to sweep the SLP, the Diagnostics Device is equipped with a Pi-Plates DAQC2Plate data acquisition board (Figure 9). The board has four {12-bit resolution analog outputs} with a voltage range from 0 to 4.096V.

It is also able to receive positive and negative analog voltage up to 12V with a 16-bit resolution. This board eliminates the need to extra amplifiers to generate the voltage sweep, and is powered directly from the Raspberry Pi, eliminating the need for extra power sources connected to the device.

\begin{figure}
  \centering
  \includegraphics[width=\columnwidth]{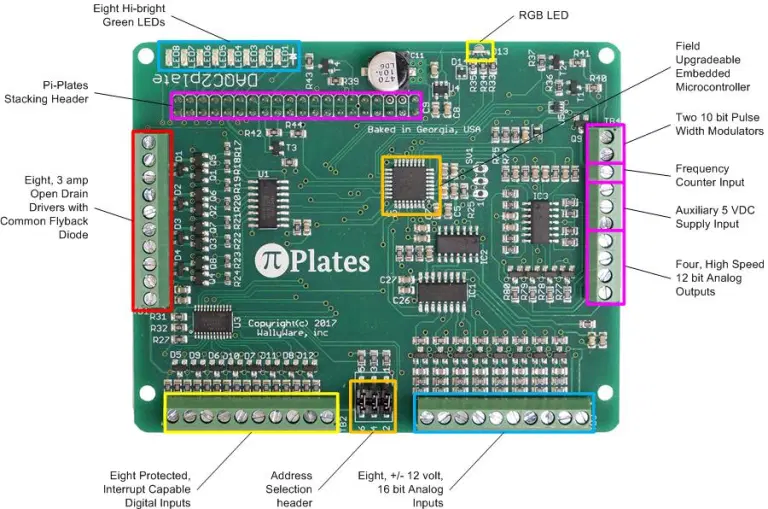}
  \caption{DAQC2Plate Data Acquisition Board\\
  https://pi-plates.com/daqc2r1/}\label{DAQC2Plate}
\end{figure}

\subsection{Amplification Circuits}
\subsubsection{Buffer Amplifier}
A Buffer amplifier was needed to prevent a loading effect on the Raspberry Pi’s output, since the Raspberry Pi is not able of provide the output current required by the circuit. The circuit of the buffer amplifier as shown in Figure 3, is done by simply connecting the feedback with a 1k ohm resistor thus the gain is unity.  An Analog Devices AD8541 was used to buffer the Raspberry Pi output voltage. The AD8541 is a general-purpose CMOS single rail-to-rail amplifier \cite{Patharkar2014PerformanceAO, Dubey2015CMOSBC}, \cite{tan2020design}.

\begin{figure}
  \centering
  \includegraphics[width=.8\columnwidth]{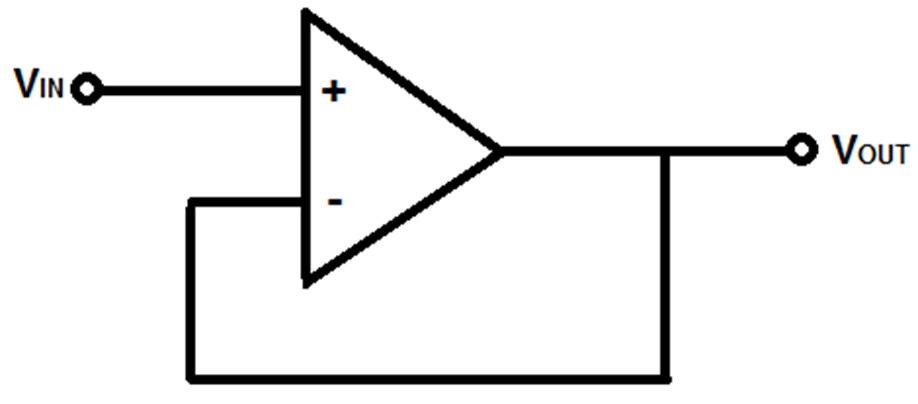}
  \caption{Buffer Amplifier \\
  http://www.learningaboutelectronics.com/Articles/Unity-gain-buffer}\label{BufferAmp}
\end{figure}
\subsubsection{Summing Amplifier}
Based on previous experiences, the design was required to generate a voltage ramp from -300V to 300V with a current rating not to exceed 20 mA. The voltage amplification of the circuit was done in stages. First, the output signal of 0-4V from the Raspberry Pi was converted to a negative and positive output swing of -2V to 2V, using a low-noise amplifier. For the input voltage of 0-4V and the available 5V Vcc, the Analog Devices OP27 is well suited. The OP27 has low skew rate, 8Mhz gain bandwidth, input voltage range of 12.3V and a maximum supply voltage of 22V.

\begin{figure}
  \centering
  \includegraphics[width=\columnwidth]{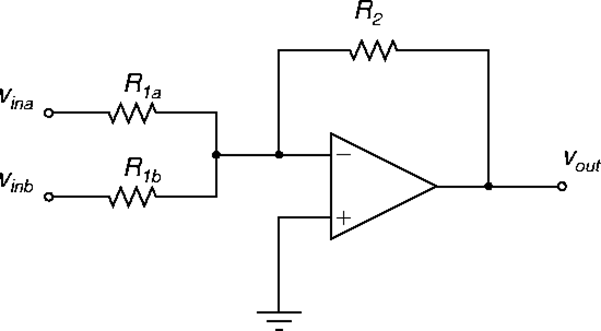}
  \caption{Buffer Amplifier \\
  http://users.cecs.anu.edu.au/\~Matthew.James/engn2211-2002/notes/ampsnode15.html}
  \label{SummingAmp}
\end{figure}

\subsubsection{High Voltage Power Operational Amplifiers}
A high voltage precision amplifier is used to amplify the -2V to 2V signal obtained from the summing amplifier, to the -300V to 300V range.
%Like the previous analysis, the goal is to obtain 300V when the input signal is -2V and -300V when the input voltage is 2V (Note the inverting polarity due to the inverting configuration).
An Apex PA95 High voltage amplifier was used to this purpose given its High Voltage supply voltage of up to 900V, low-quiescent current of 1.6 mA, and output current of 100 mA with current limiter.

\begin{figure}
  \centering
  \includegraphics[width=\columnwidth]{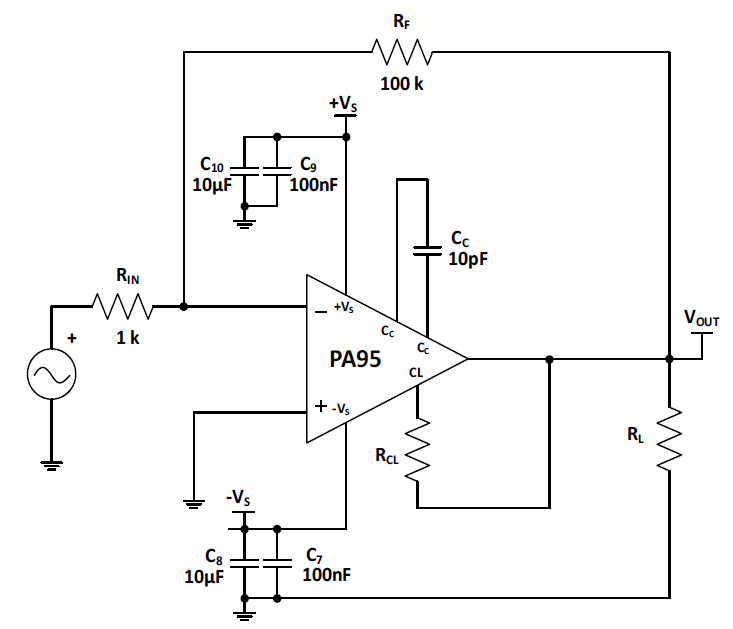}
  \caption{Inverting Amplifier}\label{PA95_typ}
\end{figure}

\subsubsection{Current Sensor Amplifier Circuit}
Since it is impossible to connect the Raspberry Pi analog input (12V max.) directly to the high voltage output to measure the current, an Analog Devices AD8479 Precision Difference Amplifier was used with a shunt resistor to obtain a voltage proportional to the current fromm the probe. The AD8479 is a very-high common-mode voltage precision difference amplifier with ±600 V common-mode voltage range, low offset voltage, low offset voltage drift, low gain drift, low common-mode rejection drift, and excellent common-mode rejection ratio (CMRR) over a wide frequency range.

%\subsection{Buck-Boost Converter}
%
%To provide the at least 350V/18mA source for the High Voltage Amplifier, a BuckBoost Converter was used. A DC-DC-8-32V-to-45-390V High Voltage Boost Converter ZVS Step up Booster was used, being  powered by the 15V DC output of the PT65-C power supply. The eBay High Voltage Boost Module was used for the design as it meets the requirements. The High Voltage Boost Module is capable to function with a voltage of 8~32V and 5A max

\subsection{Power Supply}
The Langmuir Probe Diagnostics Device requires various amplifier circuits which in turn need voltage sources of different values, including negative voltage. To meet the constraints of size and budget, it was crucial to find a power supply capable of having 3 outputs.

\begin{table}[h]
  \centering \renewcommand*\arraystretch{1.3}
  \caption{Power Supply Requirements}\label{PSA}
\begin{tabular}{|l|c|c|}
\hline
Components & Voltage (V) & Current\\
\hline
Boost Converter & 15 & 15mA\\
\hline
Summing Amplifier & ±5 & 6mA\\
\hline
Buffer Amplifier & ±5 & 6mA\\
\hline
Differential Amplifier & ±15 & 0.5mA\\
\hline
Operational Amplifier & ±350 & 18mA\\
\hline
Raspberry Pi & 5 & 4A\\
\hline
Raspberry Pi Camara & 5 & 4A\\
\hline
\end{tabular}

\end{table}

Using Table \ref{PSA}, the PT65C from Mean Well was selected since it met the requirements to supply power to most of the devices from the list.
PT65-C by Mean Well is capable of providing 5Vdc, 15Vdc and -15 Vdc with 7 A, 2.6 A and 0.7 A respectively with a size of only 3” by 5” and rated for 65 Watts.

However, to comply with safety measures, the Raspberry Pi and the touchscreen display are powered by a separate power supply. They are powered by the standard CanaKit 5A Raspberry Pi 4 Power Supply. The CanaKit Power supply has USB-C, includes noise filter for added stability.

\section{IMPLEMENTATION}

The prototype went through an iterative process. A first prototype was developed without enclosure and Human Machine Interface (HMI). For the second iteration, the HMI was implemented and tested. For the final prototype, the enclosure was introduced. There were some adjustments that were done to the electrical components between each iteration and a few of them will be discussed during this section.

\begin{figure}
  \centering
  \includegraphics[width=\columnwidth]{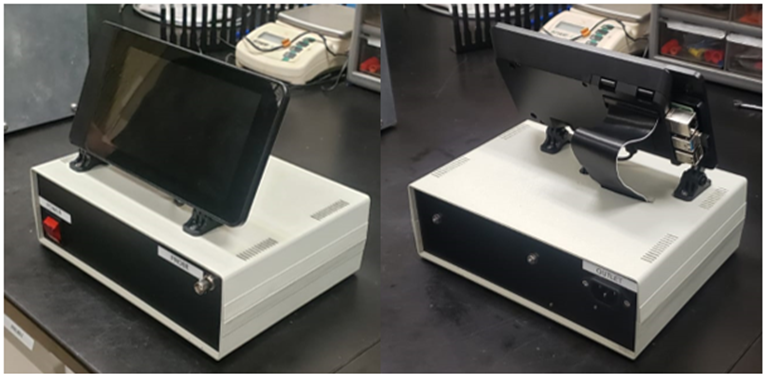}
  \caption{Final Prototype}
  \label{Prototype}
\end{figure}

\begin{figure}
  \centering
  \includegraphics[width=\columnwidth]{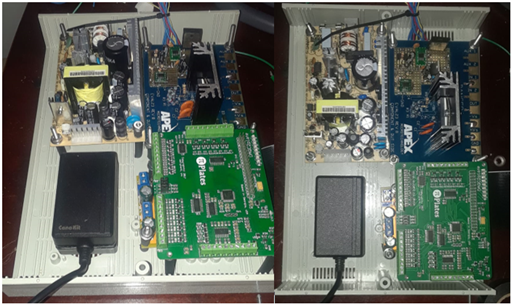}
  \caption{Final Prototype}
  \label{Prototype_inside}
\end{figure}

The final Prototype design, shown in Figures \ref{Prototype} and \ref{Prototype_inside}, presents a small compact enclosure integrating a mounting stand for the touchscreen display on the top.

The device counts with a relay block capable of interrupting the power supply, responsible to power all the electronics of the device. This reduces the power consumed by the device. The relay is triggered through the python script to only turn on the electronics of the system while generating the voltage ramp, user defined to last between 0.25 seconds to 10 seconds. This reduces the device power consumption and reduces the risk of electric shock for the user.

\subsection{Graphical User Interface}
\begin{figure}
  \centering
  \includegraphics[width=\columnwidth]{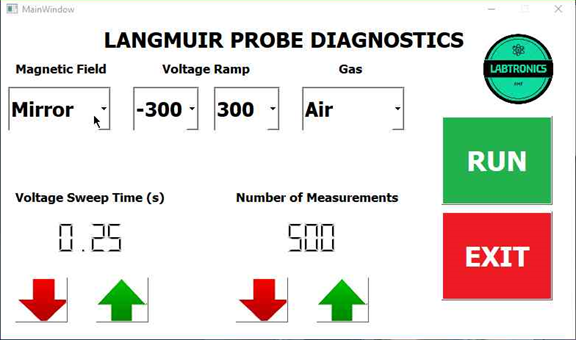}
  \caption{Graphical User Interface}
  \label{f16}
\end{figure}

The graphical user interface shown in Figure \ref{f16} was done with oversized buttons for ease of use with a touchscreen display. The graphical user interface allows to adjust the magnetic field mode to Cusp or Mirror, the SLP voltage ramp limits, gas used for producing plasma, voltage sweep time, and number of data points to be acquired. By using the Raspberry Pi’s Touchscreen Display, there is no need to use a mouse or keyboard, for a compact design.

\begin{figure}
  \centering
  \includegraphics[width=\columnwidth]{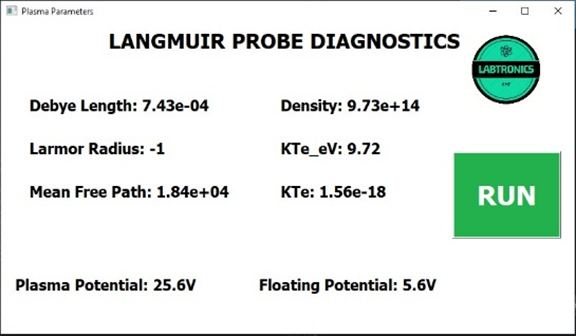}
  \caption{Real-Time Diagnostics Mode GUI}
  \label{f24}
\end{figure}

Some minor alterations were done to further improve the calculations of the plasma characteristics device. Also, a continuous measurement mode for displaying plasma characteristics in real-time was implemented (Figure \ref{f24}), by modifying the python script originally designed, by introducing some minor tweaks. The real-time diagnostics mode was successfully implemented.


\subsection{Experimental Logs}
\subsubsection{Experiment 1}

\begin{figure}
  \centering
  \includegraphics[width=\columnwidth]{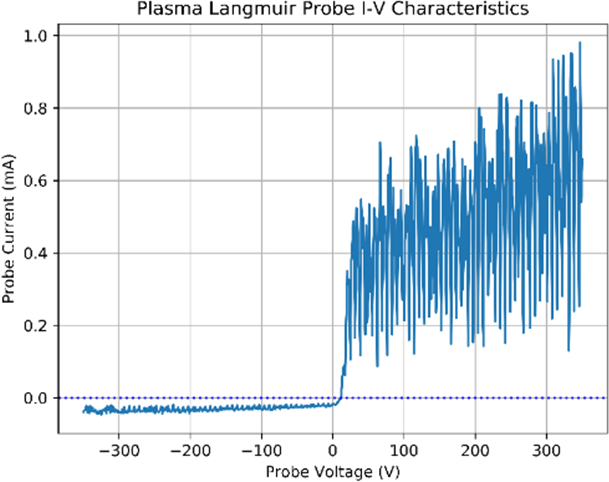}
  \caption{First Experiment IV Characteristic Raw}
  \label{f25}
\end{figure}

The first experiment consisted of only testing the python code. The code provides the user with settings, controls the generation of the voltage ramp signal, retrieve the current from the Langmuir Probe, and plots the IV characteristic from the acquired data. The IV characteristic shown in Figure \ref{f25}, was obtained in the first experiment resembling results of previous laboratory experiments.

\subsubsection{Experiment 2}
\begin{figure}
  \centering
  \includegraphics[width=\columnwidth]{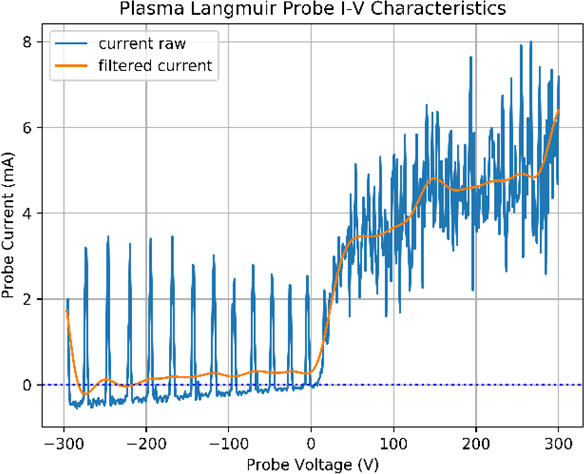}
  \caption{Second Experiment IV Characteristic}
  \label{f26}
\end{figure}

First data acquisition using hardware. During this experiment, the first prototype was used to generate a voltage ramp, read current and plot the values into an easy to interpret plot. The plot in Figure \ref{f26} shows a standard IV characteristic typically from a Langmuir Probe. However, noticeable noise is seen throughout the graph which corrupted the calculations of the plasma parameters. A grounding issue was the responsible for this behavior.

\subsubsection{Experiment 3}
\begin{figure}
  \centering
  \includegraphics[width=\columnwidth]{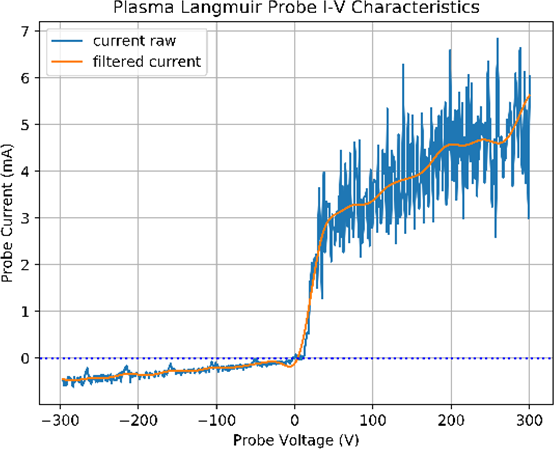}
  \caption{Third Experiment IV Characteristic}
  \label{f28}
\end{figure}

Hardware modifications significantly reduced the ripple to an acceptable value as shown in Figure \ref{f28}. Ripples apparently present due to switching action of Buck-Boost Converters. The data acquired is valid to be used for plasma diagnostics calculations. The SLP IV regions are shown in Figure 29, as displayed by the device.

\begin{figure}
  \centering
  \includegraphics[width=\columnwidth]{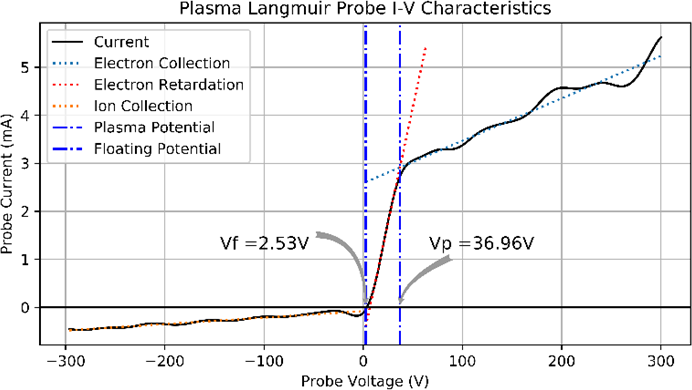}
  \caption{Third Experiment - IV Regions Plotted}
  \label{f29}
\end{figure}

\section{CONCLUSION}

The goal of this project was to design and build a plasma diagnostics device, specifically for a Single Langmuir Probe able to generate the voltage sweep required by the probe and to determine plasma parameters in a continuous fashion. Previously, this task was done by multiple devices, including a signal generator, a current sensor, and a computer capable of reading, and analyzing numerical data. This project integrated all these components into one compact, self-sufficient electronic device. The final device uses a Raspberry Pi to create a voltage signal. A series of electronic amplifiers manipulate the signal, and an electronic current sensor measures any changes in the electric current generated in the plasma. The information is retrieved by the Raspberry Pi is analyzed to calculate the different parameters of the plasma, like electron temperature, density, floating and plasma potential, mean free path, Larmor radius, and Debye Length. This information is displayed on a small touchscreen and all this process is controlled by the user using the same touchscreen as I/O device to the Raspberry Pi.
The device was teste under various plasma conditions and produces measurements consistent with other instruments in the laboratory.

%\bibliographystyle{ieeetrans}
%\bibliography{RESEARCHBIB}
% Generated by IEEEtranS.bst, version: 1.14 (2015/08/26)

\end{document}